\documentclass{emulateapj}
\usepackage{amssymb}
\usepackage{amsmath}
\usepackage{epstopdf}
\usepackage{natbib}
\usepackage[colorlinks, citecolor=blue]{hyperref}
\usepackage{hyperref}
\usepackage[all]{hypcap}

\begin{document}

\title{The Most Luminous Supernova ASASSN-15lh: Signature of a Newborn Rapidly-Rotating Strange Quark Star}

\author{Z. G. Dai\altaffilmark{1,2}, S. Q. Wang\altaffilmark{1,2}, J. S. Wang\altaffilmark{1,2}, L. J. Wang\altaffilmark{3}, and Y. W. Yu\altaffilmark{4}}

\affil{\altaffilmark{1}School of Astronomy and Space Science, Nanjing University, Nanjing 210093, China; dzg@nju.edu.cn}
\affil{\altaffilmark{2}Key Laboratory of Modern Astronomy and Astrophysics (Nanjing University), Ministry of Education, China}
\affil{\altaffilmark{3}Key Laboratory of Space Astronomy and Technology, National Astronomical Observatories, Chinese Academy of Sciences, Beijing 100012, China}
\affil{\altaffilmark{4}Institute of Astrophysics, Central China Normal University, Wuhan 430079, China}

\begin{abstract}
In this paper we show that the most luminous supernova discovered very recently, ASASSN-15lh, could have been powered by a newborn ultra-strongly-magnetized pulsar, which initially rotates near the Kepler limit. We find that if this pulsar is a neutron star, its rotational energy could be quickly lost as a result of gravitational-radiation-driven r-mode instability; if it is a strange quark star, however, this instability is highly suppressed due to a large bulk viscosity associated with the nonleptonic weak interaction among quarks and thus most of its rotational energy could be extracted to drive ASASSN-15lh. Therefore, we conclude that such an ultra-energetic supernova provides a possible signature for the birth of a strange quark star.
\end{abstract}


\section{Introduction}
\label{sec:Intro}

Fast developing sky survey programs have gathered some superluminous supernovae (SLSNe) \citep{Qui2011,Cho2011,Gal2012}
whose peak magnitudes are $\lesssim -21$ mag \citep[for a recent review on observational properties of SLSNe see][]{Gal2012}.
SLSNe have been divided into types I (hydrogen-deficient) and II (hydrogen-rich), most of which
 cannot be explained by the $^{56}$Ni-powered model \citep{Ins2013,Nic2013,Nic2014}.
 Currently, most of SLSNe-II are explained by the ejecta-circumstellar medium interaction model \citep{Che1982,Gin2012}, while
 the most prevailing model explaining SLSNe-I is the magnetar (ultra-highly-magnetized pulsar) model \citep{Ins2013,Kas2010,Woo2010},
which supposes that a nascent magnetar just after the
 core-collapse supernova converts its rotational energy to the heating energy of supernova (SN)
 ejecta\footnote{Nonrelativistic pulsar-powered supernovae were originally proposed by \citet{Ost1971}.
 Ultra-relativistic pulsar-powered shocks were put forward to predict
 temporal plateaus of gamma-ray burst afterglows \citep{Dai1998a,Dai1998b,Zhang2001,Dai2004}. This prediction is well consistent
 with the shallow decay phase (where the flux density decays as $\propto t^{-\alpha_f}$ with slope of $\alpha_f\sim 0-0.5$) of many early afterglows
 discovered by the {\em Swift} satellite \citep{Zhang2007,Yu2007,Dall2011}.}.
The magnetar-powered model has been used to explain a few SLSNe-I \citep[e.g.,][]{Ins2013,Nic2013,How2013,McC2014,Vre2014,Nic2014,Wang2015b,Kasen2015}
 as well as luminous SNe Ic with peak magnitudes $\sim -20$ mag \citep{Wang2015c,Gre2015}.

Recently, \citet{Dong2015} observed and analyzed a very luminous optical transient at redshift $z=0.2326$, ASASSN-15lh, and found that
it reached $\sim2.2\times10^{45}$ ergs~s$^{-1}$ at time $\sim15$~days after the peak. Several explanations of this unusual transient
except for the SLSN origin are implausible.
First, the authors excluded a gravitational lensing event since the observed redshift is so low that any significant contribution from
gravitational lensing is highly impossible.
Second, the active galactic nucleus origin is disfavored because the observed flux
variability and spectral slope of a normal active galactic nucleus
 are inconsistent with those of ASASSN-15lh. Blazars and related ``jetted'' sources are also excluded because of
 their featureless, non-thermal, power law spectra \citep{Dong2015}.
Third, a nuclear outburst in an active Seyfert galaxy or an event of a supermassive black hole tidally disrupting a star
is hydrogen-rich, while ASASSN-15lh's spectrum
is hydrogen-deficient. Thus, neither a nuclear outburst nor a tidal disruption event can explain ASASSN-15lh \citep{Dong2015}.
In addition, ASASSN-15lh's spectrum closely resembles another SLSN-I, PTF10cwr/SN 2010gx.
Therefore, \citet{Dong2015} concluded that ASASSN-15lh is a SLSN-I, which is much brighter than all of the other SLSNe discovered so far.
\citet{Dong2015} argued that if this SLSN is powered by $^{56}$Ni,
the required mass of $^{56}$Ni is $\gtrsim30~M_\odot$, which is enigmatical for any stellar explosion;
if the magnetar model is employed, the stellar initial rotation period could be $\sim1$\,ms.
A similar estimate for the the initial period of the magnetar was also obtained by \citet{Met2015} and \citet{Kashi2015}.
Considering the effect of gravitational radiation, \citet{Dong2015} further argued that ASASSN-15lh challenges the magnetar model.

Provided that it is a newborn neutron star (NS), this magnetar is indeed spun quickly down to a period $\gtrsim 5$~ms \citep{Mad1998}
due to a significant rotational energy loss caused by the r-mode instability \citep{And1998,Fri1998}.
Thus, the magnetar model seems to be ruled out by ASASSN-15lh.
In this paper, we attempt to resolve this difficulty and
constrain the physical nature of a magnetar assumed to power such a SN.
We find that the central engine of this SN could be a strange quark star (SQS).

It was recognized more than forty years ago \citep{Bod1971} that strange quark matter consisting almost entirely of comparable numbers of
deconfined up, down and strange quarks could be absolutely stable. If this is true, then SQSs composed of such matter
could in principle exist \citep{Wit1984} \citep[for an early simple discussion see][]{Itoh1970}. SQSs were suggested to form possibly during some astrophysical processes, e.g. SN explosions, accretion in X-ray binaries, and mergers of two NSs \citep{Gen1993,Dai1995,Cheng1996,Dai1998b}. Typically, SQSs are understood based on the phenomenological MIT bag model of strange quark matter \citep{Far1984}, but their structural and cooling features are similar to those of NSs for stellar masses above one solar mass \citep{Hae1986,Alc1986}. Thus, it is not easy to identify a SQS from observations on stellar structures and surface temperatures \citep[for a review see][]{Web2005}. Fortunately, a quick rotational energy loss due to the gravitational radiation-driven r-mode instability in a newborn (very hot) rapidly rotating NS is absent in a SQS counterpart \citep{Mad1998}. Therefore, only a newborn SQS can reach the Keplerian rotation limit.

This paper is organized as follows. In section \ref{sec:ana}, we give an upper limit
 of the initial period and corresponding magnetic field strength of a magnetar assumed to power ASASSN-15lh.
 In section \ref{sec:fit}, we fit the light curve of this SN.
In section \ref{sec:r-mode}, we analyze the r-mode instability in the magnetar and then constrain the stellar nature.
Finally, we present conclusions and discussions in section \ref{sec:dis}.

\section{Limits on magnetar parameters}
\label{sec:ana}

We start by discussing the upper limit of the initial period and the field strength
of a magnetar (with mass of $M=M_{1.4}\times1.4M_\odot$ and radius of $R=R_6\times 10^6$\,cm) assumed to power ASASSN-15lh.
The stellar luminosity due to magnetic dipole radiation is \citep{Ost1971,Sha1983}
\begin{equation}
L_{\rm mag}(t)=\frac{E_{\rm rot}}{\tau_{\rm mag}}\frac{1}{(1+t/\tau_{\rm mag})^2},
\label{equ:input-mag}
\end{equation}
where $E_{\rm rot}=(1/2)I(2\pi/P_{\rm i})^2$ is the initial rotational energy with $I$ being the moment of inertia and $P_{\rm i}$ being the initial rotation period, and $\tau_{\rm mag}$ is the spin-down timescale of the magnetar and can be expressed by
\begin{equation}
\tau_{\rm mag}=\frac{3c^3IP_{\rm i}^2}{2\pi^2B^2R^6}=4.75I_{45}P_{\rm i,ms}^2B_{14}^{-2}R_{6}^{-6}~{\rm days},\label{equ:tau-mag}
\end{equation}
where $c$ is the speed of light, $I_{45}=I/10^{45}$~g~cm$^2$, $P_{\rm i,ms}=P_{\rm i}/1\,{\rm ms}$, and $B=10^{14}B_{14}$~G is
 the surface magnetic field strength. In equation (\ref{equ:tau-mag}), it has been assumed that the angle between the magnetic axis and rotation axis is 45$^\circ$.

According to the Arnett law \citep{Arn1979,Arn1982}, which
indicates that the peak luminosity $L_{\rm pk}$ of a SN at peak time $t_{\rm pk}$ is equal to the instantaneous energy deposition rate, that is,
$L_{\rm mag}(t)|_{t=t_{\rm pk}}=L_{\rm pk}$, we obtain
\begin{equation}
P_{\rm i}^2=\frac{2{\pi}^2BR^3}{(3c^3)^{1/2}L_{\rm pk}^{1/2}}-\frac{2{\pi}^2B^2R^6t_{\rm pk}}{3c^3I}.
\label{equ:input-mag3}
\end{equation}
From equation (\ref{equ:input-mag3}), we can see that the initial period $P_{\rm i}$ reaches a maximum at
\begin{equation}
B=B_{\rm cr}\equiv\frac{(3c^3)^{1/2}I}{2R^3L_{\rm pk}^{1/2}t_{\rm pk}}
=52.1\times 10^{14}I_{45}R_{6}^{-3}L_{\rm pk,44}^{-1/2}t_{\rm pk,d}^{-1}\,{\rm G},
\label{equ:bcri2}
\end{equation}
where $L_{\rm pk,44}=L_{\rm pk}/10^{44}{\rm erg}~{\rm s}^{-1}$, and $t_{\rm pk,d}=t_{\rm pk}/1\,{\rm day}$.
The maximum value of $P_{\rm i}$ is
\begin{equation}
P_{\rm i,max}=\frac{\pi I^{1/2}}{\sqrt{2}L_{\rm pk}^{1/2}t_{\rm pk}^{1/2}}=23.9I_{45}^{1/2} L_{\rm pk,44}^{-1/2}t_{\rm pk,d}^{-1/2}\,{\rm ms}.
\label{equ:input-mag4}
\end{equation}

Based on the V-Band observation, we can see that
$t_{\rm pk}=t_{\rm pk,obs}/(1+z)$ ($z=0.2326$) is $\sim30~{\rm to}~40$~days.
Around this peak time, the two observational luminosities derived from the optical band are
$2.85_{-0.97}^{+1.67}\times10^{45}$ ergs~s$^{-1}$ and
$2.11_{-0.72}^{+1.24}\times10^{45}$ ergs~s$^{-1}$, and at $\sim15$ days after the peak time, the luminosity is
$2.16_{-0.19}^{+0.21}\times10^{45}$ ergs~s$^{-1}$ \citep{Dong2015}. Therefore,
it is reasonable to adopt $L_{\rm pk}\sim(2.0~{\rm to}~4.0)\times10^{45}$ ergs~s$^{-1}$.

The realistic stellar moment of inertia for $M>M_\odot$ is estimated by $I\simeq 0.237MR^2(1+2.84\eta+18.9\eta^4)$,
where $\eta=GM/(Rc^2)=0.208M_{1.4}R_6^{-1}$ \citep{Lat2005}. Letting $M_{1.4}=1$ and $R_{6}=1$ and substituting
these typical values into equations (\ref{equ:bcri2}) and (\ref{equ:input-mag4}), we get $B_{\rm cr}\simeq3\times 10^{13}$~G and
$P_{\rm i,max}\simeq0.76$~ms. Therefore, we conclude that ASASSN-15lh could have been powered by
an ultra-strongly-magnetized pulsar rotating with a sub-millisecond period.

The period of a rotating pulsar has a theoretical lower limit, viz., the Kepler period $P_{\rm K}$. According to \citet{Hae2009},
$P_{\rm K}\simeq C_{\rm K}M_{1.4}^{-1/2}R_6^{3/2}~{\rm ms}$, where $C_{\rm K}=0.78$ for a NS and $C_{\rm K}=0.73$ for a SQS.
Thus, it is easy to find that $P_{\rm K}\simeq 0.78$~ms for a NS and $P_{\rm K}\simeq 0.73$~ms for a SQS, provided that $M_{1.4}=R_6=1$.
These values of the Kepler period basically satisfies the following physical requirement: $P_{\rm K}\lesssim P_{\rm i,max}$.
This in fact leads to a new constraint on the stellar mass-radius relation (for $M>M_\odot$),
which can be fitted by $M\gtrsim(0.497+1.043R_6-0.113R_6^2)M_\odot$ for a NS and $M\gtrsim(0.477+0.980R_6-0.106R_6^2)M_\odot$ for a SQS.
These inequalities rule out very soft equations of state for dense matter.

\begin{figure}
\begin{center}
   \includegraphics[width=0.45\textwidth,angle=0]{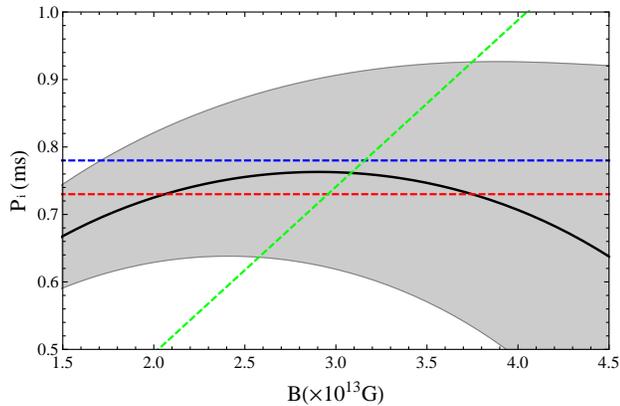}
   \caption{Constraining $P_{\rm i}$ and $B$ from observations on ASASSN-15lh for NSs
   and SQSs, assuming $M_{1.4}=1$ and $R_6=1$.
   The thick solid line and shadow region indicate the best and $1\sigma$ values of $P_{\rm i}$, respectively.
   The blue and red dashed lines are the Kepler limits of NSs and SQSs, respectively. The green dashed line is
   plotted by the condition of $\tau_{\rm mag}\gtrsim \tau_m$, which leads to $P_{\rm i,ms}\gtrsim0.247B_{13}$ (see the text in section \ref{sec:fit}).}
   \label{fig:limits}
\end{center}
\end{figure}

If the mass and radius of a central rotating pulsar are given, we can obtain constraints on the initial rotation period
and magnetic field strength from observations on ASASSN-15lh. According to equation (\ref{equ:input-mag3}), Fig. \ref{fig:limits}
shows such constraints based on $L_{\rm pk}\sim (3.0\pm1.0)\times 10^{45}\,{\rm ergs}\,{\rm s}^{-1}$ and
$t_{\rm pk}\sim (35\pm 5)\,{\rm days}$.
As a reasonable approximation, therefore, we adopt $P_{\rm i}\sim 0.8$~ms, and $B\sim B_{\rm cr}\simeq3\times 10^{13}$~G,
which is significantly weaker than the field strengths of magnetars used to power the other SLSNe-I, where the latter
are generally extremely-strongly magnetized, $B\gtrsim10^{14}$~G \citep{Kas2010,Ins2013,Nic2013,Nic2014}.

\section{Fit to ASASSN-15lh}
\label{sec:fit}

We consider a widely-adopted semi-analytical model to fit the light curve of ASASSN-15lh.
Based on the Arnett law \citep{Arn1979,Arn1982}, letting the initial radius of the progenitor
 $R_{0}\rightarrow0$ and taking into account the gamma-ray and X-ray leakage,
the luminosity of a SN can be given by
\begin{eqnarray}
L(t)=\frac{2}{\tau_{m}}e^{-\frac{t^{2}}{\tau_{m}^{2}}}
\int_0^te^{\frac{t'^{2}}{\tau_{m}^{2}}}\frac{t'}{\tau_{m}}L_{\rm mag}(t')\left[1-e^{-\tau_{\gamma}(t')}\right]dt',
\label{equ:lum}
\end{eqnarray}
where $\tau_{\gamma}(t)=At^{-2}=\left(3\kappa_{\gamma}M_{\rm ej}/4{\pi}v_{\rm sc}^2\right)t^{-2}$ is the optical depth
to gamma-rays \citep{Cha2012}, $\tau_{m}=\left({2\kappa M_{\rm ej}}/{\xi{c}v_{\rm sc}}\right)^{1/2}$ is the diffusion timescale,
$\kappa_{\gamma}$ is the opacity to gamma-rays, $\kappa$ is the opacity to optical photons,
$M_{\rm ej}$ is the ejecta mass, $v_{\rm sc}$ is the scale velocity of the ejecta \citep{Arn1982}, and $\xi\simeq 13.8$.
In fact, $v_{\rm sc}$ is approximately equal to the photospheric expansion velocity $v_{\rm ph}$, which
is estimated to be $\simeq 2.0\times 10^{4}$~km~s$^{-1}$ for ASASSN-15lh \citep{Met2015}.

\begin{figure}
\begin{center}
   \includegraphics[width=0.54\textwidth,angle=0]{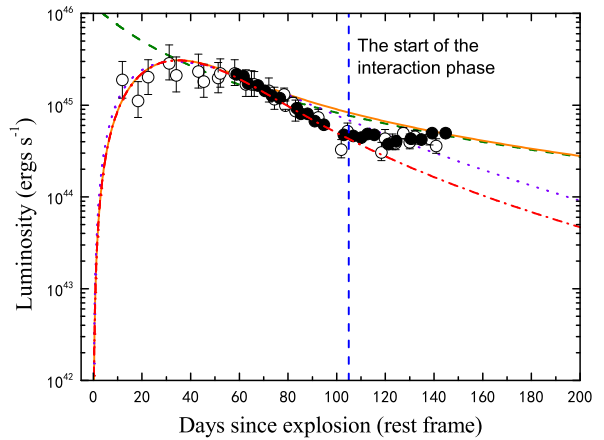}
   \caption{Fit to ASASSN-15lh using the magnetar- and $^{56}$Ni-powered models.
   Data points are taken from \citet{Dong2015}, where filled dots are bolometric luminosities derived from the optical-ultraviolet band
   and unfilled dots are derived from the optical band. The green dashed, orange red solid, red dot-dash, and purple dotted lines are
   the magnetar input curve, magnetar-powered light curve with $\kappa_\gamma\rightarrow\infty$, magnetar-powered light curve with
   $\kappa_\gamma=0.03\,{\rm cm}^2\,{\rm g}^{-1}$, and $^{56}$Ni-powered light curve with $\kappa_\gamma=0.027\,{\rm cm}^2\,{\rm g}^{-1}$,
   respectively. The late-time rebrightening is speculated to result possibly from
   an interaction between the SN ejecta and circumstellar medium, which is similar to SLSN-I iPTF13ehe \citep{Yan2015,Wang2015d}.
   The vertical blue dashed line represents the start of this interaction phase.}
   \label{fig:fit}
\end{center}
\end{figure}

We fit the observed data of ASASSN-15lh using equation (\ref{equ:lum}) and adopting $M_{1.4}=1$ and $R_{6}=1$ in Fig. \ref{fig:fit}.
The other parameters are: $v_{\rm sc}\simeq2.0\times10^{4}$~km~s$^{-1}$, $\kappa=0.1$~cm$^2$~g$^{-1}$,
$M_{\rm ej}=15M_{\odot}$ \footnote{The optical opacity $\kappa$ of the C+O-dominant ejecta is uncertain
but taken to be in the range from $0.05$ to $0.2$ cm$^2$~g$^{-1}$ in the literature. The diffusion timescale $\tau_m$ of a SN
depends on $\kappa$, the ejecta mass, and the photospheric velocity. If the photospheric velocity is fixed,
the diffusion timescale depends on the values of $\kappa$ and ejecta mass.
When $\kappa=0.1$~cm$^2$~g$^{-1}$, the observed light curve requires that the ejecta mass is frozen to $\sim 15M_{\odot}$.
 To maintain the same peak luminosity and shape of a light curve, ${\kappa}\times M_{\rm ej}={\rm constant}$ must be required.
 Hence, a smaller (larger) $\kappa$ requires a larger (smaller) $M_{\rm ej}$, i.e., the ejecta mass inferred
 from the light curve depends on the value of $\kappa$ adopted.}, $B=3.1\times10^{13}$~G, $P_{\rm i}=0.77$~ms, and $\kappa_{\gamma}=0.03$~cm$^2$~g$^{-1}$.
In this fit, the peak luminosity and the rise time are $3.04\times10^{45}$ ergs~s$^{-1}$ and $\sim 35$~days, respectively.
The value of $\tau_m$ is $\simeq 31$~days, slightly less than the rise time. Theoretically, it is required that
$\tau_{\rm mag}\gtrsim \tau_m$ for magnetar-powered SN ejecta to produce sufficiently strong photospheric emission \citep{Wang2015a},
which leads to a new constraint on the initial rotation period and magnetic field strength,
$P_{\rm i,ms}\gtrsim0.247B_{13}$. This constraint has been shown by the green dashed line in Fig. \ref{fig:limits}.
The values of $B$ and $P_{\rm i}$ adopted in Fig. \ref{fig:fit} are in agreement with those estimated above.

Replacing the magnetar input function by the $^{56}$Ni decay input function \citep[see equation 4 in][]{Wang2015c},
we can reproduce the light curve powered by the $^{56}$Ni cascade decay, shown in Fig. \ref{fig:fit}. For the $^{56}$Ni-powered SNe Ic,
the fiducial value\footnote{For the C+O ejecta (SNe Ia and Ic),
the opacity to gamma-rays $\kappa_{\gamma}$ is assumed to be gray and does rely on the energy of gamma-rays $E_{\gamma}$ and $Y_e$
(where $Y_e\sim0.5$ is the number of electrons per baryon). Therefore it is reasonable to assume the same $\kappa_{\gamma}$
for all SNe Ia and Ic with different masses. Monte Carlo simulations performed
 by \citet{Swa1995} showed that the value of $\kappa_{\gamma}$ is $(0.06\pm0.01)Y_e$ cm$^2$~g$^{-1}$.
When $E_{\gamma} > 0.1$ MeV, the value of $\kappa_{\gamma}$ depends only on the value of $Y_e$, i.e.,
$\kappa_{\gamma} \simeq 0.05 Y_e \simeq 0.025-0.03$ cm$^2$~g$^{-1}$ \citep[see Fig. 1 in][]{Swa1995}.
Because $E_{\gamma}$ ($^{56}$Ni$\rightarrow$$^{56}$Co)~=~0.148~MeV and $E_{\gamma}$ ($^{56}$Co$\rightarrow$$^{56}$Fe)~=~0.847~MeV,
we have $\kappa_{\gamma}~\sim~0.025-0.03$ cm$^2$~g$^{-1}$. Thus, $\kappa_\gamma\sim0.027$~cm$^2$~g$^{-1}$ was adopted
in the literature investigating SNe Ia and Ic.
For complete gamma-ray trapping, the total $^{56}$Ni mass is the same as in the case of $\kappa_{\gamma}= 0.027$~cm$^2$~g$^{-1}$, because
in the case of a larger $\kappa_\gamma$, the gamma-ray leakage only influences the post-peak luminosity.
 In the case of no gamma-ray trapping, however, any amount of $^{56}$Ni does not provide an optical-IR luminosity for SNe,
 since no gamma-ray is trapped and converted to optical-IR radiation.} of $\kappa_{\gamma}$ is usually assumed to be 0.027 cm$^2$~g$^{-1}$ \citep[e.g.,][]{Cap1997,Maz2000,Mae2003}.
The required mass of $^{56}$Ni is $\sim260M_\odot$, significantly larger than both the lower limit ($\sim30M_\odot$)
estimated in \citet{Dong2015} and the mass of $^{56}$Ni synthesized by any possible stellar explosion,
demonstrating the invalidity of the $^{56}$Ni-powered model.
We speculate that the rebrightening in the late-time light curve \citep{Bro2015,Dong2015}
could be due to an interaction between the SN ejecta and circumstellar medium (where the start of this interaction phase
is marked by the vertical blue dashed line in Fig. \ref{fig:fit}), which is similar to iPTF13ehe
\citep{Yan2015,Wang2015d,Dong2015b}. Thus, fitting the late-time bump is beyond the scope of this paper and we no longer discuss it here.

Since $L_{\rm pk}{\propto}IP_{\rm i,max}^{-2}t_{\rm pk}^{-1}$ in equation (\ref{equ:input-mag4}),
 if $\tau_{m}$ ($\sim t_{\rm pk}$) is varied, $L_{\rm pk}$ should
 also change. Decreasing $\tau_{m}$ ($\sim t_{\rm pk}$) would result in
 an increase of $L_{\rm pk}$, but the magnetar input curve intersects
 every peak luminosity. For example,
when $\tau_{m}$ is halved, the peak luminosity must be $\sim6\times10^{45}$ ergs~s$^{-1}$,
about one magnitude brighter than the peak luminosity of ASASSN-15lh ($\sim 3\times10^{45}$ ergs~s$^{-1}$).
Therefore, it is expected to observe SNe brighter than this SN in the future.

\section{R-mode instability analysis}
\label{sec:r-mode}

In a rotating compact fluid star, the r-mode instability is excited by gravitational radiation
and suppressed by viscosities \citep{And1998,Fri1998,Owen1998}.
For a NS, the typical driving timescale of gravitational radiation is $t_{\rm gw}\simeq47M_{1.4}^{-1}R_6^{-4}P^6_{\rm ms}$\,s,
where $P_{\rm ms}=2\pi\times10^3/\Omega$ with $\Omega$ being the rotation angle velocity,
the dissipation timescale due to shear viscosity is $t_{\rm sv}\simeq6.7\times10^{7}M_{1.4}^{-5/4}R_6^{23/4}T_9^2$\,s,
and $t_{\rm bv}\simeq2.7\times10^{11}M_{1.4}R_6^{-1}T_9^{-6}P_{\rm ms}^2$\,s due to bulk viscosity,
where $T_9$ is the core temperature in units of $10^9\,$K \citep{Owen1998}.

For a SQS, the timescale of gravitational radiation is $t_{\rm gw}\simeq21M_{1.4}^{-1}R_6^{-4}P^6_{\rm ms}$\,s, and the dissipation timescale due to shear viscosity is $t_{\rm sv}\simeq7.4\times10^{7}\bar{\alpha}_{\rm s,0.1}^{5/3}M_{1.4}^{-5/9}R_6^{11/3}T_9^{5/3}$~s, where $\bar{\alpha}_{\rm s,0.1}$ is the strong coupling constant in units of $0.1$ \citep{Mad2000}. The
dissipation timescale $t_{\rm bv}$ due to bulk viscosity as a result of the non-leptonic weak interaction among quarks in a SQS is highly dependent on temperature. The importance of this dissipation mechanism was first recognized by \citet{Wang1984}. For $T\ll10^9$ K,
$t_{\rm bv}\simeq7.9M_{1.4}^2R_6^{-4}T_9^{-2}P_{\rm ms}^2m_{100}^{-4}$ s, and for $T\gg10^9$ K,
$t_{\rm bv}\simeq2.3R_6^2T_9^{2}P_{\rm ms}^4m_{100}^{-4}$ s \citep{Mad2000}, where $m_{100}$ is
the strange quark mass in units of 100\,MeV.

An additional viscosity is due to the effect of a boundary layer under the solid crust of an old NS
\citep{Bil2000}. For a newborn (very hot) NS, however, this effect may be unimportant, because the stellar outer layers are fluidic or at most in a mixed fluid-solid state within the first $\sim 10^7$\,s \citep{Lin2000}. In this early stage, any superconductivity and/or superfluidity phase in the stellar interior does not occur and thus its effect can also be neglected.

We now study the effect of the r-mode instability on the spin evolution of newborn NSs and SQSs.
Following \citet{Ho2000}, \citet{And2001}, and \citet{Yu2009a}, we parameterize the evolution of spin ($\Omega$) and r-mode amplitude ($\alpha$) as
\begin{equation}
\frac{d\Omega }{dt} =
-\frac{3 \alpha ^2 \tilde{J} \Omega }{A_+}
\left(\frac{1}{t_{\rm sv}}+\frac{1}{t_{\rm bv}}\right)
+\frac{N_{\rm mag}}{A_+ M R^2},\label{equ:oevo}
\end{equation}
and
\begin{equation}
\frac{d\alpha }{dt}=\frac{\alpha }{t_{\rm gw}}
-\frac{\alpha A_-}{A_+}\left(\frac{1}{t_{\rm sv}}+\frac{1}{t_{\rm bv}}\right)
-\frac{\alpha  N_{\rm mag}}{A_+MR^2\Omega},\label{equ:aevo}
\end{equation}
where
$A_\pm=\tilde{I}\pm 3 \alpha ^2 \tilde{J}/2$
with $\tilde{I}=0.261$ and $\tilde{J}=1.635\times10^{-2}$.
The torque of magnetic dipole radiation can be
expressed by $N_{\rm mag}=-B^2R^6\Omega^3/(12c^3)$, where the inclination angle of the magnetic axis to the rotation axis has been taken to be $45^\circ$. As it grows, the r-mode amplitude would eventually saturate, at which $(d\alpha/dt)|_{\alpha=\alpha_s}=0$ with $\alpha_s$ being the saturation amplitude. After then, $\alpha_s$ keeps nearly constant and the spin evolution follows
\citep{And2001}
\begin{equation}
\frac{d\Omega }{dt}=-\frac{3 \alpha_s ^2 \tilde{J} \Omega }{A_-(\alpha_s)}
\frac{1}{t_{\rm gw}}+\frac{N_{\rm mag}}{M R^2 A_-(\alpha_s)}\label{equ:oevos}.
\end{equation}
The cooling processes in the magnetar must be known to solve equations (\ref{equ:oevo}) and (\ref{equ:aevo}).
As usual, we consider the direct URCA processes to describe the cooling of a SQS and the modified URCA processes
in a NS as well as the viscous heating within these two classes of stars.

\begin{figure}
\begin{center}
   \includegraphics[width=0.45\textwidth,angle=0]{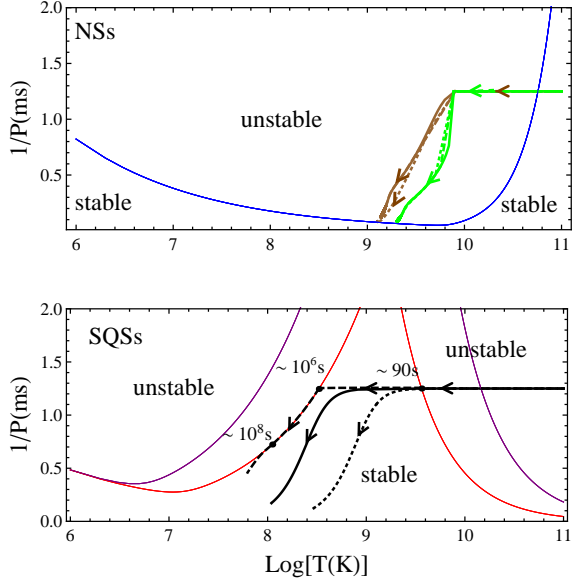}
   \caption{Rotation frequency versus temperature for the r-mode instability: NSs ({\em upper panel})
   and SQSs ({\em lower panel}) for $M_{1.4}=1$ and $R_6=1$.
   The blue, purple and red lines are critical curves between r-mode stability and instability regions.
   The thick dashed, solid, and dotted lines show magnetic fields of $3\times10^{12}$\,G, $3\times10^{13}$\,G,
   and $3\times10^{14}$\,G, respectively. For NSs, the green lines correspond to $\alpha_s=1$
   and the brown lines to $\alpha_s=0.1$. For SQSs, the purple line corresponds to $m_{100}=2$,
   and the red line to $m_{100}=1$, where the thick dashed line gives $\alpha_s\simeq 0.01$ after $\sim 10^6$\,s.
   The numbers near black dots are times since the birth of a SQS. The arrows represent evolution directions.}
   \label{fig:r-mode}
\end{center}
\end{figure}

For a newborn rapidly-rotating NS, the r-mode amplitude is usually assumed to grow to a constant $\alpha_s$
of order unity \citep{Owen1998}. This order is supported by numerical simulations \citep{Ster2001,Lin2001}.
Recent studies \citep{Yu2009b,Alf2012} show that $\alpha_s$ is in the range of $\sim 0.1$ to a few.
%which would be expected to be testable with r-mode gravitational waves \citep{Myt2015}.
Therefore, it is reasonable to take $\alpha_s=0.1$ and 1. In addition, the initial amplitude $\alpha_0=10^{-6}$.
The spin and temperature evolution can be calculated and shown in Fig. \ref{fig:r-mode}.
We see from the upper panel of this figure that a newborn NS rotating with $\sim 0.8$ ms quickly reaches the window of instability,
then the r-mode amplitude grows to $\alpha_s$ within hundreds of seconds, and finally this NS keeps r-mode
unstable until it departs from the instability window at period $P\sim5$\,ms. The {\em green lines} (and {\em brown lines})
with different magnetic fields are overlapped, indicating that gravitational radiation dominates the spin evolution.

For a SQS, the instability window splits into a high-temperature part and a low-temperature part
due to a large bulk viscosity associated with the nonleptonic weak interaction among quarks \citep{Mad2000}.
We see from the lower panel of Fig. \ref{fig:r-mode}
that a newborn SQS with $P_{\rm i}\sim 0.8$\,ms initially lies in the high-temperature instability region,
but this instability window does not affect significantly the spin evolution, because the star cools so fast
that the r-mode instability has no enough time ($\sim 90$\,s) to spin it down obviously.
Therefore, when it exits from the high-temperature instability window, the star has a period close to the initial one. Subsequently, stars with different magnetic fields have
different spin evolution curves. First, a star with field of $\sim3\times 10^{12}$\,G ({\em thick dashed line}), when it cools to a few $10^8$ K,
reaches the low-temperature instability window, at which time ($\sim 10^6$\,s) the star almost keeps its initial period,
and then it starts to be spun down significantly due to the r-mode instability. About $10^8$\,s later
after the birth, the star departs from the low-temperature instability window and loses its angular momentum via magnetic dipole radiation.
Thus, such a star cannot power ASASSN-15lh because of a low $L_{\rm mag}$ and r-mode instability occurring $\sim 10^6$\,s later after the birth.
Second, after it leaves the high-temperature instability window, a star with field of $\sim3\times 10^{13}$\,G ({\em thick solid line})
is spun down quickly by magnetic dipole radiation, so that it does never meet the low-temperature instability window. Such a SQS, if its
initial period $P_{\rm i}\sim 0.8$\,ms, could just power ASASSN-15lh. Third, a star with much stronger field
(e.g., $\sim3\times10^{14}$\,G, indicated by the {\em thick dotted line}) loses its rotational energy so quickly
via magnetic dipole radiation that it can only drive a SN (with a similar peak time) much fainter than ASASSN-15lh,
as shown in equation (\ref{equ:input-mag3}).

\begin{figure}
\begin{center}
   \includegraphics[width=0.45\textwidth,angle=0]{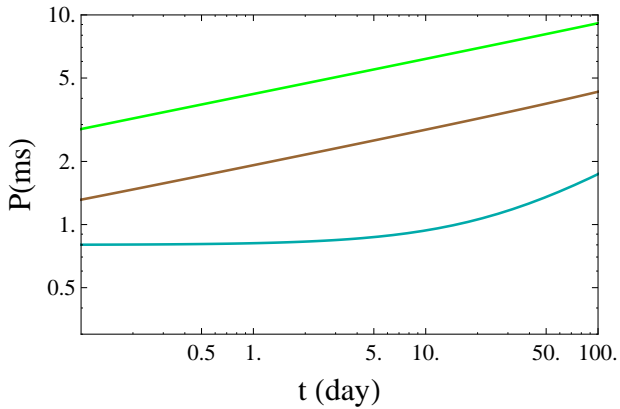}
   \caption{Evolution of the rotation period for NSs with $M_{1.4}=1$, $R_6=1$ and $B=3\times10^{13}$\,G.
   The cyan line is obtained without r-mode instability.
   The green and brown lines are plotted for $\alpha_s=1$ and
   $0.1$, respectively. All of the lines have considered the effect of magnetic dipole radiation.}
   \label{fig:pt}
\end{center}
\end{figure}

We also calculate the contribution of the gravitational radiation-driven r-mode instability to spin evolution of a NS
in detail, which is shown in Fig. \ref{fig:pt}. We can see that for a NS, gravitational radiation will spin down the magnetar
significantly and bring away most of the stellar rotational energy if $\alpha_s$ is in the range of 0.1 to unity.
This implies that a nascent NS seems to be unable to drive ASASSN-15lh.

\section{Conclusions and discussions}
\label{sec:dis}

ASASSN-15lh is the most luminous SN discovered so far \citep{Dong2015}.
 We have shown that if this SN was powered by the $^{56}$Ni cascade decay, $\sim260M_\odot$ of
 $^{56}$Ni must be synthesized, far exceeding the $^{56}$Ni yield produced by any
 single explosion of a massive star. Therefore, the $^{56}$Ni-powered model is excluded.

 We alternatively adopted the magnetar-powered model to explain the high luminosity of this SN.
We found that the initial period of the magnetar powering ASASSN-15lh is
 $\sim P_{\rm i,max}\sim 0.8$ ms, which is very close to the Kepler limit. The magnetar's field is only
 $B\sim 3\times 10^{13}$~G. Detailed calculations indicate that a newborn NS with these stellar parameters is subject to
 the r-mode instability and loses most of the rotational energy via r-mode gravitational radiation, but
a nascent SQS counterpart does not suffer from this instability. Hence, ASASSN-15lh could have been powered by a SQS rather than a NS.
In addition, we showed that SQSs with similar periods but with much weaker or much stronger magnetic fields are implausible to explain this SN.

It should be noted that many other SLSNe-I have been assumed to be powered by magnetars with periods less than $5$~ms,
 e.g. SN~2011ke ($P_{\rm i}\simeq 1.7$~ms), SN~2011kf ($P_{\rm i}\simeq 2.0$~ms), SN~2010gx ($P_{\rm i}\simeq 2.0$~ms) \citep{Ins2013},
 PTF 12dam ($P_{\rm i}\simeq 2.6$~ms) \citep{Nic2013}, SN~2013dg ($P_{\rm i}\simeq 2.5$~ms), LSQ12dlf ($P_{\rm i}\simeq 1.9$~ms), SSS120810 ($P_{\rm i}\simeq 1.2$~ms) \citep{Nic2014}.
If these magnetars are NSs, they could quickly lose their rotational energy via r-mode gravitational radiation.
Thus, while the NS-magnetar model is still challenged in explaining these SLSNe-I,
SQSs as their central engines might be able to resolve this question.
Therefore, an accurate estimate of the initial rotation periods of newborn magnetars powering SLSNe-I
would help us distinguish between NSs and SQSs, and could eventually provide new insights into
the stellar nature.

Although the initial period of the magnetar powering ASASSN-15lh is approximately equal to the
Kepler limit, the theoretical upper limit of a SN peak luminosity has not yet been reached.
We would expect that future survey programs are promising to discover SLSNe luminous than
ASASSN-15lh \footnote{\citet{Dong2015} estimated that the rate of ASASSN-15lh-like events is
$\sim 0.28-3.7$ Gpc$^{-3}$ yr$^{-1}$ (at the 90\% confidence level), significantly lower than
the rate of SLSNe-I (i.e., $\sim 11-152$ Gpc$^{-3}$ yr$^{-1}$). Thus, ASASSN-15lh-like events are very rare.
Perhaps SLSNe that are more luminous than ASASSN-15lh have been missed in previous SN surveys. The cadence of
 future survey programs might be higher than that of current survey programs
  and therefore might detect SNe more luminous than ASASSN-15lh.}.

Finally, what we would point out is that soft equations to state for some exotic matter such as kaon condensation
in NSs seem to be unlikely, because the maximum mass of stars containing such matter is significantly lower than
the mass measurements ($M_{\rm pulsar}\simeq 2.0M_\odot$) of two pulsars \citep{Demo2010,Anto2013}. Even if kaons occur in the core region of a neutron star, their effect on the r-modes could be insignificant. This is because the outer regions where the r-modes are mainly located do not include kaons, so that we needn't consider a bulk viscosity associated with kaons. On the other hand, SQSs discussed here are not ruled out
if the equation of state for strange quark matter is stiff enough \citep{Alf2007}.

\acknowledgments
We thank the referee for his/her very helpful comments and suggestions that have allowed us to improve our manuscript.
We also thank Subo Dong, Yizhong Fan, Naoki Itoh, Kazumi Kashiyama, Xiangyu Wang, Xuefeng Wu, and Bing Zhang for their useful comments.
This work was supported by the National Basic Research Program (``973" Program)
of China (grant No. 2014CB845800) and the National Natural Science Foundation of China (grant Nos. 11573014 and 11473008).

%\clearpage

\end{document}